\documentclass[a4paper,11pt]{article} 
\usepackage[margin=0.5in]{geometry} 
\usepackage{amssymb}
\usepackage{amsmath,epsfig, epsf, epic, graphicx} 
\usepackage{fancybox, lscape, subfigure} 
\usepackage{bm}
\usepackage{natbib}
\usepackage{float}
\usepackage[margin=2cm]{caption}
\usepackage{setspace}

\newcommand\independent{\protect\mathpalette{\protect\independenT}{\perp}}
\def\independenT#1#2{\mathrel{\rlap{$#1#2$}\mkern2mu{#1#2}}}

\title{Doubly robust matching estimators for high dimensional confounding adjustment}
\author{Joseph Antonelli, Matthew Cefalu\footnote{Joseph Antonelli and Matthew Cefalu are co-lead authors}, Nathan Palmer, Denis Agniel}

\begin{document}
\date{}
\maketitle{}

\doublespacing

\abstract{Valid estimation of treatment effects from observational data requires proper control of confounding. If the number of covariates is large relative to the number of observations, then controlling for all available covariates is infeasible. In cases where a sparsity condition holds, variable selection or penalization can reduce the dimension of the covariate space in a manner that allows for valid estimation of treatment effects. In this article, we propose matching on both the estimated propensity score and the estimated prognostic scores when the number of covariates is large relative to the number of observations. We derive asymptotic results for the matching estimator and show that it is doubly robust, in the sense that only one of the two score models need be correct to obtain a consistent estimator. We show via simulation its effectiveness in controlling for confounding and highlight its potential to address nonlinear confounding. Finally, we apply the proposed procedure to analyze the effect of gender on prescription opioid use using insurance claims data.}

\section{Introduction}
The goal of many research studies is to estimate the effect of a treatment on an outcome. If the treatment is not randomized, as is the case in observational studies, then care must be taken to ensure valid estimation. One common issue faced is the selection of covariates. The omission of a single confounding variable may lead to biased inference, while inclusion of unnecessary covariates may inflate the variance. This issue becomes more pertinent as the number of covariates increases, and standard methods for confounding adjustment will fail if the number of covariates is large compared to the sample size.

Recommendations for covariate selection when estimating treatment effects are varied but can be loosely classified into three categories: (1) control for all observed covariates; (2) selection based on substantive knowledge \citep{vanderweele2011new}; and (3) data-driven approaches \citep{van2010collaborative,de2011covariate,vansteelandt2012model,wang2012bayesian,zigler2014uncertainty,wilson2014confounder}. In the age of big data, where access to and use of electronic medical records, administrative databases, and large-scale genomic and imaging datasets is increasingly common, the number of covariates available for analysis continues to grow. When the number of covariates is large relative to the number of observations, controlling for all observed covariates becomes infeasible and selection based on substantive knowledge becomes impractical. As such, the focus of this paper will be on data-driven approaches for dimension reduction. 

A variety of methods allow the inclusion of a high-dimensional vector of covariates in regression given that a sparsity condition holds. Arguably the most popular, the lasso \citep{tibshirani1996regression} places a penalty on the absolute value of the coefficients for the covariates, which forces many of the coefficients to zero, leading to a more parsimonious model. Many similar penalization methods have been proposed \cite[e.g.]{fan2001variable,zou2005regularization,zou2006adaptive}. These approaches suffer in the context of effect estimation as they are designed for prediction. Estimation and prediction are very different statistical challenges, and the variables that one requires for valid estimation may be different than those needed for prediction. Fitting a lasso model for the outcome focuses only on the relationship of each variable with the outcome and ignores any association with the treatment. Such a method will tend to omit variables that are strongly associated with treatment but weakly associated with the outcome. Omission of these variables may lead to bias in the estimated effect.

Other methods have been developed to address this issue by performing variable selection or model averaging aimed at selection of confounders for use in effect estimation.  \cite{wang2012bayesian} proposed a Bayesian model averaging procedure that uses an informative prior to place more weight a priori on outcome models that include covariates associated with the exposure. Many ideas have built on this prior specification to address the issue of confounder selection and model uncertainty \citep{talbot2015bayesian, wang2015accounting, cefalu2013model}. There also exists a small literature on dimension-preserving statistics that can be used in a similar manner as propensity scores to balance confounders between levels of a binary treatment. These approaches can be found in \cite{ghosh2011propensity, nelson2013information, lue2015inverse} and the references within. All of the aforementioned approaches have been shown to work well in identifying confounders or adjusting for confounding. However, none of these approaches can handle a high-dimensional vector of confounders. 

Recent literature has focused on the scenario in which the number of confounders may exceed the number of observations. \cite{wilson2014confounder} took a decision theoretic approach to confounder selection and showed that this approach had strong connections to the adaptive lasso, but with weights designed to select confounders instead of predictors. \cite{belloni2013inference} and \cite{farrell2015robust} utilized standard lasso models on both the exposure and outcome, identifying confounders as variables that enter into either lasso model, then fitting an unpenalized regression model or doubly robust estimator on the reduced set of covariates. \cite{ertefaie2015variable} addressed the issue of selecting weak confounders in small sample sizes by penalizing a joint likelihood on the exposure and outcome. Regularization for effect estimation is adopted from a Bayesian perspective in \cite{hahn2016bayesian} by reparameterizing the likelihood and using horseshoe priors on the regression coefficients. \cite{ghosh2015penalized} utilized penalization in the potential outcomes framework, though their goal is to identify covariates that modify treatment effects. 

The approach proposed in this paper, called {\em doubly robust matching}, aims to handle high-dimensional confounding by matching on both the propensity score \citep{rosenbaum1983central} and the prognostic score \citep{hansen2008prognostic}. Recent work by \cite{leacy2014joint} has shown that matching on both scores in low-dimensional settings can lead to improved inference over simply matching on the propensity score. We extend these ideas to higher dimensions by incorporating penalization into both the propensity score model and the prognostic score model. In addition, we demonstrate that matching on both scores simultaneously is doubly robust, in the sense that the treatment effect is consistently estimated if either the propensity score or the prognostic score is correctly specified. Using high-dimensional simulations, we show that our doubly robust matching estimator is superior to other doubly robust estimators because it is not sensitive to extreme propensity scores and it appears to be robust to misspecification of both scores.


\section{Methodology and asymptotic results}
\label{sec:methods}

\subsection{Notation and framework}
\label{sec:notation}
Suppose we have collected $N$ independent observations from $(Y,W,X)$, where $Y$ is the observed outcome, $W$ is a binary treatment of interest, and $X$ is a $P$-dimensional vector of covariates such that $P$ may be larger than $N$. Let $Y(1)$ be the potential outcome under treatment and let $Y(0)$ be the potential outcome under control \citep{rubin1974estimating}. Our goal is the estimation of the average treatment effect defined as:
\begin{align}
	\tau = E[Y(1) - Y(0)],
\end{align}
\noindent where the expectation is over the population of interest. 

For identification of the average treatment effect, we rely on the stable unit treatment value assumption (SUTVA), strong ignorability, and positivity. SUTVA is described elsewhere \citep{little2000causal}, but it can be understood as two conditions: the treatment received by one observation or unit does not affect the outcomes of other units and the potential outcomes are well-defined in the sense that there are not different versions of the treatment that lead to different potential outcomes. Strong ignorability and positivity are defined as: \\
\indent \textit{Strong Ignorability:} $Y(1), Y(0) \independent W \vert X$

\textit{Positivity:} $0 < \varphi(X) < 1$ for all $X$\\
\noindent where $\varphi(X)=P(W=1 \vert X)$ denotes the propensity score \citep{rosenbaum1983central}. Under these assumptions, the average treatment effect is identified conditional on the propensity score:
\begin{align*}
	\tau = E[ E\{Y | W=1 , \varphi(X) \} - E\{ Y | W=0 , \varphi(X) \} ].
\end{align*}

In addition, we define prognostic scores for each potential outcome as any scores $\Psi_0(X)$ and $\Psi_1(X)$ that satisfy the following conditions \citep{hansen2008prognostic}:
\begin{align}
	Y(0) &\independent X ~\vert~ \Psi_0(X) \label{prog_score}\\
    Y(1) &\independent X ~\vert~ \Psi_1(X) \label{prog_score1}
\end{align}
For brevity, we restrict our attention to the case of no effect modification so that there is a single prognostic score, $\Psi(X)$, that satisfies (\ref{prog_score}) and (\ref{prog_score1}). Under these assumptions, the average treatment effect is identified conditional on the prognostic score:
\begin{align*}
	\tau = E[ E\{Y | W=1 , \Psi(X) \} - E\{ Y | W=0 , \Psi(X) \} ].
\end{align*}

We consider a specific specification of a prognostic score where $\Psi(X) = E[Y(0)|X]$, although other prognostic scores are possible. For a discussion of the implications of effect modification, see Section \ref{sec:effectmod}. 








\subsection{Identifiability and double robustness}

In this section, we show that the average treatment effect is identified when conditioning on both the propensity score and the prognostic score. Interestingly, identifiability is maintained even if one of the two scores is incorrectly specified. This can be interpreted as a double robustness property, in which only one of the propensity score and prognostic score must be correctly specified to identify the average treatment effect. 

\noindent\textit{{\bfseries Theorem 1:} Assume that SUTVA, strong ignorability, and positivity hold. Further, assume there is no effect modification. Let $\varphi(X)$ be the true propensity score, let $\Psi(X)$ be a true prognostic score, and let $h(X)$ be any arbitrary function of $X$. Then,
\[ Y(1),Y(0) \independent W ~\vert~ \varphi(X) , h(X)~~~{\rm{and~~~}} Y(1),Y(0) \independent W ~\vert~ \Psi(X) , h(X).\]
}
\noindent A proof of Theorem 1 can be found in Web Appendix A. Theorem 1 states that the treatment assignment is ignorable conditional on a correctly specified propensity or prognostic score and any other arbitrary function of the covariates. This result allows identification of the average treatment effect as:
\[\tau = E\left[ E\left\{Y | \varphi(X), h(X), W = 1\right\} - E\left\{Y | \varphi(X), h(X), W = 0\right\} \right],\]
\noindent or
\[\tau = E\left[E\left\{ Y | \Psi(X), h(X), W = 1\right\} - E\left\{Y | \Psi(X), h(X), W = 0\right\}\right].\]


Theorem 1 motivates and provides very strong justification for matching on both the propensity score and the prognostic score, as only one of the two scores must be correct to obtain valid estimates of the average treatment effect. Formal double robustness (i.e. consistency) based on matching is shown in Section \ref{sectheory}. These results theoretically verify the simulation study by \cite{leacy2014joint}, which showed that bias remains small even when one of the scores is misspecified. 

When the propensity score is misspecified, we leverage the assumption of no effect modification to identify the average treatment effect through the prognostic score. If there is effect modification, identifiability can be achieved by conditioning on two prognostic scores, one for each potential outcome, or by targeting the average treatment effect on the treated. For further discussion, see Section \ref{sec:effectmod} and the Supplementary Materials.  

\subsection{Definition of the doubly robust matching estimator}
\label{sec:def}

Following \citet{leacy2014joint}, we propose to estimate the average treatment effect by matching on both the propensity score and the prognostic score. As indicated in Theorem 1, matching on both scores is doubly robust. For this reason, we will call our estimator the \text{doubly robust matching estimator} (DRME). 

\noindent The DRME takes the form:
\begin{align}
	\tau(\theta) = \frac{1}{N} \sum_{i=1}^N (2W_i - 1) \left( Y_i - \frac{1}{M} \sum_{j \in \mathcal{J}_M(i, \theta)}Y_j \right), \label{eqn:tau}
\end{align}
\noindent where $M$ is the number of matches for each subject, $\mathcal{J}_M(i, \theta)$ is the set of matches for subject $i$, and $\theta$ is a set of parameters for the score models, adapting the notation from \citet{abadie2006large}. Intuitively, the DRME finds $M$ matches for each subject based on the similarity of their score values. The mean of the $M$ matches is used to estimate the unobserved potential outcome for each subject, and the overall estimate of the average treatment effect is the mean difference between the potential outcomes for the $N$ subjects in the data.

The DRME depends intrinsically on the propensity and prognostic score models through the matching set $\mathcal{J}_M(i, \theta)$. In the rest of this section, we will clarify this relationship by defining $\theta$ and $\mathcal{J}_M(i, \theta)$. We propose the two following models for the propensity and prognostic scores:
\begin{align}
	\varphi(X)&=P(W=1|X) = g(X'\gamma_w) \label{eqn:MainModel2} \\
    \Psi(X)&=E(Y|W=0,X) = f(X'\gamma_y) \label{eqn:MainModel},
\end{align}
\noindent where $f(\cdot)$ and $g(\cdot)$ are inverse link functions  and $\theta$ in \eqref{eqn:tau} is the concatenation of $\gamma_w$ amd $\gamma_y$.
Importantly, as discussed in Theorem 1, we only require that one of \eqref{eqn:MainModel2} and \eqref{eqn:MainModel} hold.  Let 
\begin{align}
\gamma^*_w &= \max_\gamma E_W[W\log\{g(X'\gamma)\} + (1-W)\log\{1-g(X'\gamma)\}]\\
\gamma^*_y &= \min_\gamma E_Y\{Y - f(X'\gamma)\}^2
\end{align}
be the possibly mis-specified targets of estimation.

Standard methods can be used to estimate (\ref{eqn:MainModel2}) and (\ref{eqn:MainModel}) when $P \ll N$. However, when the covariate space is high-dimensional, it is very challenging to perform estimation and inference without additional assumptions. To this end, we assume \textit{sparsity}: that the true number of target parameters (and thus the true number of covariates required for valid effect estimation) is much smaller than the total number of observations.  Letting $\| \cdot \|_0$ denote the number of nonzero elements of a vector, we define sparsity in our context as $\label{eqn:sparsity} \| \gamma_y^* \|_0 \leq s$ and $\| \gamma_w^* \|_0 \leq s$, where $s$ is an integer satisfying $s \ll N$. For more details regarding sparsity and its effect on high-dimensional estimation, as well as conditions on the covariate design matrix $X$, see \citep{bickel2009simultaneous,negahban2009unified,van2008high}. We note that the assumption of sparsity may be relaxed to allow $s$ to depend on $N$ with the consequence of having to adjust rates of convergence to depend on $s$. 

For high-dimensional $P$, we propose to estimate (\ref{eqn:MainModel2}) and (\ref{eqn:MainModel}) using lasso models. Let
\begin{align}
\widehat{\gamma}_y &= \underset{\gamma}{\operatorname{argmin}} \sum_{i:W_i = 0} (Y_i - f(X_i'\gamma))^2 + \lambda_y \sum_{j=1}^P|\gamma_j| \label{eqn:lasso_y} \\
\widehat{\gamma}_w &= \underset{\gamma}{\operatorname{argmax}} \sum_{i=1}^n \left[W_i\log\{g(X_i'\gamma)\} + (1-W_i)\log\{1-g(X_i'\gamma)\}\right] + \lambda_w \sum_{j=1}^P|\gamma_j|, \label{eqn:lasso_w}
\end{align}
where $\lambda_y$ and $\lambda_w$ may be chosen to agree with asymptotic results or via cross-validation. Then the estimated propensity score and prognostic score are given by $\widehat{\varphi}(X)= g(X'\widehat{\gamma}_w)$ and $\widehat{\Psi}(X)= f(X'\widehat{\gamma}_y)$, respectively. Letting $Z = [\widehat{\varphi}(X), \widehat{\Psi}(X)]$, we define the matching set as: 
\begin{align*}
	\mathcal{J}_M(i, \widehat{\theta}) = \left\{ j=1,...,N: W_j = 1 - W_i,  \left(\sum_{k:W_k = 1 - W_i} I(||Z_i - Z_k|| < ||Z_i - Z_j||) \right) \leq M \right\}.
\end{align*}
\noindent In practice, calipers are frequently used to ensure good matches and to ensure no matches outside of the common support. This is well known to change the quantity being estimated, and a nice discussion of this can be found in \cite{iacus2015theory}, but it can help in small samples to reduce bias of the matching estimator. Asymptotically, fixed-width calipers do not drop observations when positivity holds; therefore, we will not incorporate calipers in our asymptotic results. We will utilize calipers in our simulation study and in the analysis of insurance claims data found in Section \ref{sec:aetna}.

\subsection{Consistency of the doubly robust matching estimator}\label{sectheory}

In this section, we demonstrate the consistency of the DRME for estimating the average treatment effect and show that it is doubly robust, in the sense that the average treatment effect is consistently estimated when either the model for the propensity score or the model for the prognostic score is correctly specified. We do not require that both are correctly specified. These results hold for high dimensions at a rate no slower than the standard rate for high-dimensional estimators, again, when either of the two models is correctly specified. \\
%
%
\textit{{\bfseries Theorem 2:} Assuming SUTVA, strong ignorability, positivity, no effect modification, the regularity conditions necessary for asymptotic consistency of the lasso, sparsity as defined in (\ref{eqn:sparsity}), that at least 1 of the 2 high dimensional models is correctly specified, and additional weak conditions on the distribution of the data available in the Web Appendix, then}
\begin{align*}
	\tau(\widehat{\theta}) - \tau = O_p\left(\sqrt{\frac{\log P}{N}}\right).
\end{align*}
Theorem 2 has several important implications. First, matching on both a high-dimensional propensity score and a high-dimensional prognostic score is consistent. 
Second, this consistency only requires one of the two models to be correctly specified. This is the matching version of the well known double robustness property from the inverse weighting literature \citep{bang2005doubly}. Third, this result directly implies that in low-dimensional settings we have root-$N$ consistency and double robustness.  \\
\indent \textit{Sketch Proof of Theorem 2:} First, note that we can write the error of our estimator as:
\begin{align} \label{eqn:decomp}
	\tau(\widehat{\theta}) - \tau &= \left(\tau(\widehat{\theta}) - \tau(\widetilde{\theta})\right) + \left( \tau(\widetilde{\theta}) - \tau\right) 
\end{align}
where $\widetilde{\theta} = (\gamma^*_w, \gamma^*_y)$ denotes the probability limit of $\widehat{\theta}$. It is important to note that $\widetilde{\theta}$ here does not necessarily represent the true propensity and prognostic score parameters. The first component of (\ref{eqn:decomp}) is the error that arises from needing to estimate the parameters of the two score models, while the second component is the error induced by the matching process. 

We examine the asymptotic behavior of each component separately and details can be found in Web Appendix B. Loosely speaking, the first component is no slower than the $\sqrt{\frac{\log P}{N}}$ rate inherited from estimating the parameters of the lasso models and does not require correct specification of either model. The second component requires at least one of the score models to be correctly specified and has the $N^{-1/2}$ rate from matching on two scores, which follows directly from \cite{abadie2006large}. Combining the rates of convergence from the two components gives the final result.

\subsection{Implications of effect modification}
\label{sec:effectmod}

The results of the prior sections were derived under the assumption of no effect modification. If we relax this assumption, then conditioning on a single prognostic score $\Psi(X)$ is no longer sufficient for the identification of the average treatment effect. Instead, a separate prognostic score is needed for each potential outcome as defined in (\ref{prog_score}) and (\ref{prog_score1}). Theorem 1 is easily relaxed to allow effect modification by conditioning on both of these prognostic scores, $\Psi_0(X)$ and $\Psi_1(X)$, in addition to the propensity score. A proof of this result is provided in Web Appendix A. However, the rate of convergence in Theorem 2 may potentially suffer when matching on more than two scores because, in general, matching on more than two scores is consistent at a rate slower than root-N \citep{abadie2006large}. 

An alternative to matching on multiple prognostic scores in the presence of effect modification is to target the average treatment effect on the treated (ATT). The ATT is defined as $E[Y(1) - Y(0)~|~ W=1]$. Interestingly, the prognostic score for the potential outcome under control, $\Psi_0(X)$, defined in (\ref{prog_score}) is sufficient for identification of the ATT. This implies that both Theorem 1 and Theorem 2 hold for the ATT when matching on the propensity score and the prognostic score under control. Thus, regardless of the presence or absence of effect modification, our results show that matching on an estimated propensity and prognostic score in high-dimensions is consistent for the ATT with rate no slower than $\sqrt{\frac{\log P}{N}}$.

\subsection{Estimation of standard errors}
\label{sec:se}
Measuring the uncertainty of the doubly robust matching estimator is difficult due to the high-dimensional nature of the models used to estimate the propensity and prognostic scores. Limiting distributions for estimators based on propensity score matching have only recently been developed \citep{abadie2016matching}, and the estimation of uncertainty around lasso estimates is an ongoing topic of research. Combining the two to provide a limiting distribution from which inference can be performed is a difficult task and a topic of further research. Here, we provide an approximation to the standard error and assess its ability to provide valid confidence intervals through simulation. Conditional on the matches, any matching estimator can be written as a weighted average of the observed data:
\begin{align}
\widehat{\tau} = \frac{\sum_{i=1}^n W_i R_i Y_i}{\sum_{i=1}^n W_i R_i} - \frac{\sum_{i=1}^n (1 - W_i) R_i Y_i}{\sum_{i=1}^n (1 - W_i) R_i},
\end{align}
\noindent where $R_i$ is the weight given to subject $i$ in the estimator. In the case of the doubly robust matching estimator described in Section \ref{sec:def}, $R_i = 1+\frac{K_i}{M}$, where $K_i$ is the number of times subject $i$ is used as a match. Therefore, given an estimate of the residual variance, $Var(Y \vert W, X)$, which we denote $\widehat{\sigma}^2$, we can approximate the standard error as:
\begin{align}
\widehat{se}(\widehat{\tau}) = \frac{\widehat{\sigma}^2 \sum_{i=1}^n W_i R_i^2}{(\sum_{i=1}^n W_i R_i)^2} + \frac{\widehat{\sigma}^2 \sum_{i=1}^n (1 - W_i) R_i^2}{(\sum_{i=1}^n (1 - W_i) R_i)^2}. \label{eqn:se}
\end{align}

\noindent For this paper we will estimate the residual variance by fitting a lasso model to the outcome regressed against the treatment and covariates and taking the average squared residual from the fitted model. This estimate of the variance does not account for uncertainty in estimation of the two scores. Therefore, this standard error underestimates the true standard error and may lead to anti-conservative interval estimates. We assess the performance of the estimation of standard errors in Section \ref{sec:sim_se}.

\section{Simulation study}
\label{sec:sims}

We compare the DRME with several competing approaches, including standard regression models. Details of the data-generating mechanisms are left to the relevant sections, but in all cases, we simulate a binary treatment $W$, a continuous outcome $Y$, and independent standard normal covariates $X$. We consider the following set of estimation techniques:

\begin{enumerate}
	    \item Naive approach that compares the mean of $Y$ in the treated and control groups (Naive)
    \item Estimating the true model (Oracle)
    \item Fitting a lasso model to the outcome only, penalizing the potential confounders, but not penalizing the treatment effect (Outcome lasso)
    \item The double post selection approach of \cite{belloni2013inference}. This involves fitting the treatment model in Equation \ref{eqn:lasso_w} and an outcome model as in the outcome lasso approach. The union of the covariates with nonzero coefficients from these models is then used to fit a standard linear model.  (Double Post Selection)
    \item Inverse probability weighted estimator with a lasso propensity score model (lasso IPW)
    \item Doubly robust approach of \cite{farrell2015robust} that uses the same working models as the double post selection, but fits a doubly robust estimator using the resulting covariates (Farrell)
    \item Doubly robust estimator from \cite{bang2005doubly} fit with lasso models for propensity score and outcome models (lasso DR)
    \item Matching approach that matches on a high-dimensional propensity score estimated from a lasso regression of $W$ on all the covariates (Propensity Score Matching)
    \item Matching approach that matches on a high-dimensional prognostic score estimated from a lasso regression of $Y$ on all the covariates in the controls only (Prognostic Score Matching)
    \item Matching approach that matches on both the high-dimensional propensity and prognostic scores (Doubly Robust Matching)
\end{enumerate}

\noindent For all matching procedures, full matching was used with calipers of 0.5 standard deviations on each of the matching variables to ensure that no poor matches were used in the data set. Throughout, the \texttt{glmnet} R package was used to perform the lasso. All tuning parameters were chosen through cross-validation using the default arguments of the function \texttt{cv.glmnet}. 

\subsection{Linear confounding}
\label{sec:LinearSim}

First, we explore the scenario where the true treatment and outcome models are linear in the covariates on the logit and identity scales, respectively, i.e Equation \ref{eqn:MainModel2} and Equation \ref{eqn:MainModel} hold. We set $N=200$ and P=$1000$ and simulate the treatment and outcome from the following models:
\begin{align}
	W &\sim Bernoulli\left\{\frac{exp(0.4X_1 + 0.9X_2 - 0.4X_3 - 0.7X_4 - 0.3X_5 + 0.6X_6)} {1 + exp(0.4X_1 + 0.9X_2 - 0.4X_3 - 0.7X_4 - 0.3X_5 + 0.6X_6)}\right\} \\
    Y &\sim Normal(-2 + W + 0.9X_1 - 0.9X_2 + 0.2X_3 - 0.2X_4 + 0.9X_7 - 0.9X_8, \sigma^2 = 1). \label{eqn:outcome_sim}
\end{align}
\noindent Therefore, covariates 1 through 4 are confounders, and notably covariates 3 and 4 are ``weak'' confounders in the sense that they have small associations with the outcome. Covariates 5 and 6 are instruments only associated with the treatment, while covariates 7 and 8 are predictive only of the outcome. The true average treatment effect is 1, which coincides with the regression coefficient for $W$ in (\ref{eqn:outcome_sim}). 

Table \ref{tab:sim1} shows the absolute bias, standard deviation (SD), and mean squared error (MSE) from this simulation for each of the estimators. The absolute bias is calculated as the absolute difference between the mean of the 1000 estimates and the truth. Matching on the propensity score, matching on the prognostic score, outcome lasso, and lasso IPW all result in substantial bias (more than 20\%). Each of these approaches relies on a single model, which appears to be undesirable in this high-dimensional setting. The double post selection, doubly robust matching, and Farrell estimators rely on two models, increasing the chance of adjusting for the important confounders. This is verified in the simulation as all three have smaller biases (7.5\%, 8.5\% and 8.7\%) and have MSEs that compare favorably to the oracle outcome model. The double post selection approach has the smallest MSE in this setting, due to the linear relationship between the covariates and outcome. The double post selection model is fitting the correct outcome model in this case as it relies on the linearity assumption, while the doubly robust matching procedure does not (i.e. it is a nonparametric matching estimator). 


\begin{table}[ht]
\centering
\begin{tabular}{lrrr}
  \hline
type & Absolute bias & SD & MSE \\ 
  \hline
Oracle & 0.002 & 0.160 & 0.026 \\ 
  Naive & 0.490 & 0.285 & 0.321 \\ 
  Outcome Lasso & 0.290 & 0.176 & 0.115 \\ 
  Double post selection & 0.075 & 0.198 & 0.045 \\ 
  Lasso IPW & 0.365 & 0.243 & 0.192 \\ 
  Farrell & 0.087 & 0.370 & 0.145 \\ 
  Lasso DR & 0.226 & 0.169 & 0.080 \\ 
  Propensity score matching & 0.255 & 0.466 & 0.282 \\ 
  Prognostic score matching & 0.242 & 0.240 & 0.116 \\ 
  Doubly robust matching & 0.085 & 0.232 & 0.061 \\ 
   \hline
\end{tabular}
\caption{Absolute bias, standard deviation, and mean squared error from the simulation of Section \ref{sec:LinearSim} across 1000 replications.}
\label{tab:sim1}
\end{table}

\subsection{Nonlinear confounding}
\label{sec:NonlinearSim1}
It is also of interest to examine the performance of the respective approaches when either the treatment or outcome models are nonlinear functions of the confounders. In this setting, approaches that assume the covariates enter into the models linearly will be misspecified and will no longer validly estimate the causal effect of interest. We use the same simulation framework as in Section \ref{sec:LinearSim}, only now we simulate the treatment and outcome from the following models:
\begin{align}
	W &\sim Bernoulli(\frac{exp(0.3X_1^2 + 0.5X_1^3 - 0.3X_2^4 + 0.4X_3^2)} {1 + exp(0.3X_1^2 + 0.5X_1^3 - 0.3X_2^4 + 0.4X_3^2)}) \\
    Y &\sim Normal(-2 + W - 0.5X_1 + 0.5X_2^2 + 0.4X_2^3 + 0.3X_3^2, \sigma^2 = 1).
\end{align}	
\noindent The first three covariates are confounders while the rest are noise. Each of the confounders has a nonlinear association with the treatment, outcome, or both. It is important to note that all of the estimated models are the same as in Section \ref{sec:LinearSim}, which assume that the covariates enter into the systematic component of the models linearly. The lone exception is the oracle model, which again takes the form of the true regression function. 

Table \ref{tab:sim2} shows the results of the simulation across 1000 replications. We see that none of the approaches, with the exception of the true model, are able to estimate the treatment effect without bias. Importantly, the doubly robust matching estimator proposed in this paper substantially reduces the bias relative to any other approach. The bias is 6.7\% for the doubly robust matching estimator, while the next smallest bias is 25.3\% for the prognostic score matching estimator. The doubly robust matching estimator also has the lowest MSE of the non-oracle estimators at 0.133.
\begin{table}[ht]
\centering
\begin{tabular}{lrrr}
  \hline
type & Absolute bias & SD & MSE \\ 
  \hline
Oracle & 0.007 & 0.026 & 0.026 \\ 
  Naive & 0.611 & 0.296 & 0.461 \\ 
  Outcome Lasso & 0.607 & 0.272 & 0.442 \\ 
  Double post selection & 0.365 & 0.281 & 0.212 \\ 
  Lasso IPW & 0.570 & 0.287 & 0.407 \\ 
  Farrell & 0.414 & 0.350 & 0.294 \\ 
  Lasso DR & 0.553 & 0.268 & 0.378 \\ 
  Propensity score matching & 0.368 & 0.426 & 0.317 \\ 
  Prognostic score matching & 0.253 & 0.276 & 0.140 \\ 
  Doubly robust matching & 0.067 & 0.359 & 0.133 \\
   \hline
\end{tabular}
\caption{Absolute bias, standard deviation, and mean squared error from the simulation of Section \ref{sec:NonlinearSim1} across 1000 replications.}
\label{tab:sim2}
\end{table}

\subsection{Investigation of estimated standard errors}
\label{sec:sim_se}

In this section, we assess the viability of our proposed estimate of the standard error for the DRME to obtain valid inference and interval coverage. We simulated data under the same scenario as Section \ref{sec:LinearSim} and varied $N \in \{200,500,1000,2000\}$ and $P \in \{200,500,1000,2000\}$, while keeping track of 95\% interval coverage. Table \ref{tab:SE} shows the confidence interval coverage probabilities for each combination of $N$ and $P$. We see that the estimated variance from \eqref{eqn:se} generally achieves near nominal coverages, with the lone exception being when the sample size is very small and the number of covariates is very large. This is the most difficult scenario in which the uncertainty in estimation of the propensity and prognostic scores is the highest. It is important to note, however, that the doubly robust matching estimator is somewhat biased in this scenario as seen in Table \ref{tab:sim1}. This means that coverage below 95\% is not solely due to underestimation of standard errors, but also due to the bias in the estimator. We have found that if we perform a bias correction on the confidence intervals using the empirically estimated bias, then coverage increases from 89\% to 92\% when $N=200$ and $P=2000$, indicating only slightly anti-conservative standard errors.  

\begin{table}[ht]
\centering
\begin{tabular}{ | r | r | r | r | r |}
  \hline
   & P = 200 & P = 500 & P = 1000 & P = 2000 \\ 
  \hline
  N = 200 & 0.948 & 0.927 & 0.958 & 0.887 \\ 
   N = 500 & 0.965 & 0.961 & 0.954 & 0.968 \\ 
   N = 1000 & 0.957 & 0.961 & 0.964 & 0.971 \\ 
   N = 2000 & 0.950 & 0.928 & 0.964 & 0.939 \\ 
  \hline
\end{tabular}
\caption{Coverage probabilities for a variety of data dimensions using the proposed standard error estimate.}
\label{tab:SE}
\end{table}

\subsection{Sensitivity to assumptions and data generating mechanisms}

Web Appendices C-G provide a number of additional simulation results assessing the performance of the doubly robust matching estimator. We investigated scenarios with different strengths of confounding, different nonlinear data generating mechanisms, scenarios where only one of the two models is misspecified, and different sample sizes and covariate dimensions. We find that the proposed estimator performs quite well across all scenarios, particularly when both models are misspecified. When only one of the two models is misspecified, it performs competitively with any of the existing doubly robust estimators, while it greatly reduces MSE when both models are incorrect. We empirically confirmed our theoretical results by showing consistency when one of the models is misspecified, and by finding that the MSE of the estimator converges at a rate faster than $\sqrt{\frac{\log P}{N}}$. 

\section{Analysis of post-surgical prescription opioid use}
\label{sec:aetna}

In this section, we investigate the difference between males and females in the amount of opioids prescribed to commercially insured individuals after surgery. The United States is currently experiencing an epidemic of opioid dependence and abuse. If males or females were systematically prescribed more opioids, that could have significant effects on downstream addiction and could indicate an area for policy intervention. There is some controversy regarding causal estimates of immutable characteristics such as gender. While there exist studies aiming to estimate the causal effect of gender \citep{boyd2010untangling}, others have argued against this because one can not intervene or manipulate gender \citep{holland1986statistics}. \cite{greiner2011causal} argue that it is relevant to estimate the causal effect of the perception of gender, rather than gender itself, as this could hypothetically be intervened upon. Regardless as to whether causal effects of gender are well defined, we believe that it is interesting to identify differences across gender after controlling for baseline characteristics.

To investigate this question, surgeries were ascertained from a de-identified administrative database of insurance claims at Aetna, Inc., a large, national commercial managed healthcare company. This database includes all 37,651,619  million members with medical and pharmacy insurance coverage between 2008 and 2016. Data includes all medical and pharmacy claims during the study period, as well as basic demographic information. Surgeries were identified via International Classification of Disease, version 9 (ICD-9) procedure codes. Members were required to have six months of medical coverage, as well as three months of pharmacy coverage, before and after surgery. If a member had multiple surgeries that met the inclusion criteria, we only analyzed the first one. A total of $N = 205,934$ surgeries were included. 

The outcome of interest is the total days supply of opioids for which the member filled a prescription in the 90 days following surgery. Opioids were identified in the database as drugs associated with the following common primary ingredients: codeine, fentanyl, hydrocodone, hydromorphone, morphine, oxycodone, oxymorphone, or tramadol. Injected drugs were excluded. Due to a heavy right-skew in the total days supply, we log-transformed total supply. Because observed differences in opioid prescription between sexes could be due to systematic over- or under-prescription, as well as due to differences in age, surgery types, or overall health, we considered a broad range of potential confounders: surgery date, surgery type, birth year, patient relationship to insurance subscriber, and all pre-surgical diagnosis codes observed within 6 months of the surgery date. Diagnosis codes that occur in less than 50 members and diagnosis codes that occur more than four times as often in one sex than the other were excluded. In total, there were 3,696 covariates included in this analysis.

\begin{figure}[ht]
\centering
	  \includegraphics[width=0.55\linewidth]{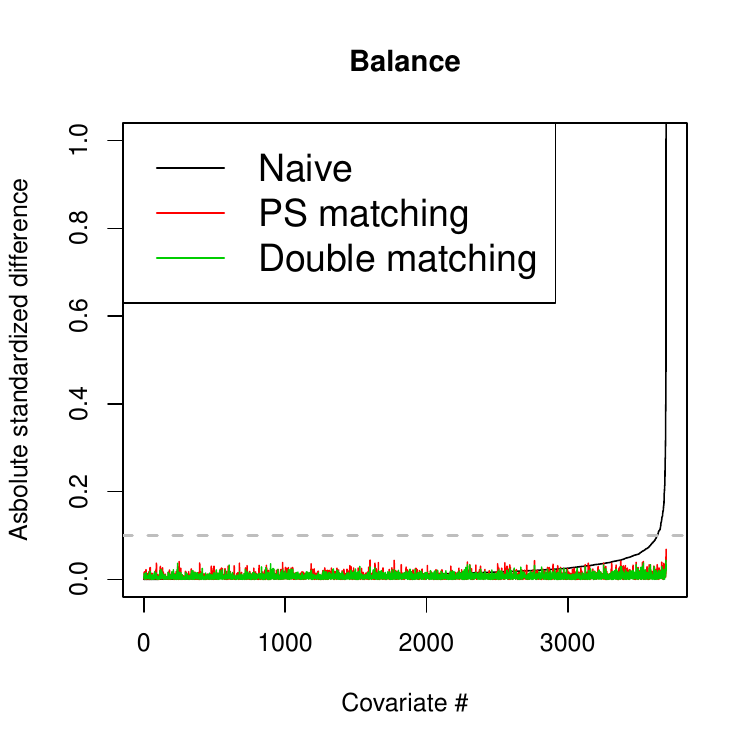}
\caption{Balance of covariates before matching, after propensity score matching, and after doubly robust matching. Balance is measured in terms of the absolute standardized difference in means between the treated and control groups. }
\label{fig:balance}
\end{figure}

One advantage of using matching-based procedures is the ability to assess balance of the covariates before and after matching by looking at the absolute standardized difference in means between the males and females for each covariate. Figure \ref{fig:balance} shows the absolute standardized difference for each covariate before matching, after propensity score matching, and after doubly robust matching. We could have also included a line for matching on the prognostic score; however, this line is very similar to the naive line as it is not intended to improve the covariate balance between males and females. It appears that both propensity score matching and doubly robust matching are achieving desirable levels of balance: the absolute standardized difference is less than 0.1 for all covariates. This indicates that both approaches are successful in removing differences between males and females with respect to all observed covariates. Table \ref{tab:balance} indicates that doubly robust matching is doing exceptionally well with regards to covariate balance, as it obtains a lower maximum and average absolute standardized difference across the covariates than propensity score matching. 

\begin{table}
\centering
\begin{tabular}{lrrr}
  \hline
&  \multicolumn{3}{c}{Absolute Standardized Difference} \\
Type & Mean & Unbalanced Mean & Maximum \\ 
  \hline
Naive & 0.02 & 0.19 & 1.91 \\ 
  Propensity score matching & 0.01 & 0.02 & 0.07 \\ 
  Doubly robust matching & 0.01 & 0.01 & 0.04 \\ 
   \hline
\end{tabular}

\caption{Illustration of the balance of the covariates before and after matching. Mean is the average absolute standardized difference across all covariates. Unbalanced mean is the same metric, except averaged only over covariates who had a naive balance greater than 0.1. Maximum is the largest absolute standardized difference across all covariates. }\label{tab:balance}
\end{table}

We estimate the difference in the log-total days supply of opioids using the same set of approaches evaluated in Section \ref{sec:sims}. The outcome lasso, lasso DR, and lasso IPW approaches do not have existing approaches for uncertainty assessment, and therefore we do not include confidence intervals for these estimators. Tuning parameters for the treatment and outcome lasso models were chosen via cross validation. In total, 467 and 1823 covariates had nonzero coefficients in the outcome and treatment models, respectively. Table \ref{tab:estimates} presents the estimated difference between males and females in this population of commercially insured patients.

\begin{table}[ht]
\centering
\begin{tabular}{lrr}
  \hline
Estimator & Difference (95\% CI) & Standard Error \\ 
  \hline
Naive & -0.055 (-0.065, -0.045) & 0.005 \\
Outcome lasso & -0.017 (--, --) & -- \\
Double Post Selection & -0.016 (-0.026, -0.006) & 0.005 \\
Lasso IPW & -0.066 (--, --) & -- \\
Farrell & -0.017 (-0.068, 0.034) & 0.026 \\
Lasso DR & -0.009 (--, --) & -- \\
  Propensity score matching & 0.006 (-0.012, 0.024) & 0.009 \\
  Prognostic score matching & 0.011 (0.001, 0.021) & 0.005 \\
  Doubly robust matching & -0.007 (-0.025, 0.010) & 0.009 \\
   \hline
\end{tabular}
\caption{Estimated difference between males and females in the log-total days supply of opioids for which members filled a prescription in the 90 days following surgery. Negative estimates indicate that males fill prescriptions with a supply of fewer days.}
\label{tab:estimates}
\end{table}

The naive approach estimates a statistically significant difference between males and females, indicating that males receive about 0.06 log-days fewer (or about 95\% of the supply of females). With the exception of the lasso IPW estimator, all of the estimators that adjust for the high-dimensional set of confounders attenuate the difference between males and females markedly. The doubly robust matching procedure estimates that males, on average and after controlling for the observed covariates, receive a supply that is for 0.007  fewer log-days, or about 99.3\% as long, and the confidence interval runs from 0.01 log-days more supply to 0.025 log-days less supply. Regardless of statistical significance, this estimated difference is not practically meaningful.

\section{Discussion}
\label{sec:discussion}

In this paper, we propose a doubly robust matching estimator for high-dimensional counfounding adjustment. This work has extended the literature on confounding adjustment in two important ways. First, we have shown theoretically that matching on a high-dimensional score such as a propensity or prognostic score is consistent with the usual rate for high-dimensional estimators, $\sqrt{\frac{\log P}{n}}$. This shows that matching presents a useful approach when trying to estimate treatment effects in difficult, high-dimensional settings. Second, matching on both the propensity score and the prognostic score is doubly robust; that is, as long as one of the two scores is correctly specified, the matching procedure is consistent. 

Many existing approaches in the literature require correct modeling of the relationship between the outcome and confounders. Models that allow for complex interactions and nonlinearities in the confounders are very difficult (if not impossible) to implement in high dimensions. Our simulations suggest that the doubly robust matching estimator is fairly robust when both the propensity score model and the prognostic score model are misspecified, indicating that simple models for the scores can be used to remove much of the bias. This extra robustness of matching estimators has been seen before as \cite{leacy2014joint} found matching estimators to be the most robust to model misspecification and \cite{waernbaum2012model} found matching estimators to be more robust than doubly robust estimators under misspecification. Our results show that these ideas extend to high-dimensional settings and further justify the use of matching estimators. 

One limitation of the doubly robust matching estimator is the current inability to derive the asymptotic variance. While this is certainly a theoretical limitation, we illustrated that reasonable confidence intervals can be constructed by ignoring the uncertainty in the estimation of the models used to create matches.

\section*{Acknowledgements}
Funding for this work was provided by National Institutes of Health (ES000002, ES024332, ES007142, ES026217 P01CA134294, R01GM111339, R35CA197449, P50MD010428)

\appendix{}

\section{Proof of Theorem 1}

We will prove a more general version of Theorem 1 that does not require the assumption of no effect modification. Using the same notation as the main text, recall that we define prognostic scores for each potential outcome as any scores $\Psi_0(X)$ and $\Psi_1(X)$ that satisfies the following conditions \citep{hansen2008prognostic}:
\begin{align*}
	Y(0) &\independent X ~\vert~ \Psi_0(X) \\
    Y(1) &\independent X ~\vert~ \Psi_1(X) 
\end{align*}

\noindent\textit{{\bfseries Theorem 1:} Assume that SUTVA, strong ignorability, and positivity hold. Let $\varphi(X)$ be the true propensity score, let $\Psi_0(X)$ and $\Psi_1(X)$ be true prognostic scores as defined above, and let $h(X)$ be any arbitrary function of $X$. Then,}
\begin{align*} 
	Y(1),Y(0) \independent W &~\vert~ \varphi(X) , h(X) \\
    Y(0) \independent W &~\vert~ \Psi_0(X), \Psi_1(X) , h(X) \\
    Y(1) \independent W &~\vert~ \Psi_0(X), \Psi_1(X) , h(X)
\end{align*}

The main elements of the proof of Theorem 1 follow directly from the following more general result. \\

\textit{Corollary 1: For any random variables $Z$ and $X$ and any functions $f$ and $g$} \[ (Z \independent X ~|~ f(X)) \Rightarrow (Z \independent X ~|~ f(X) , g(X) )\]

\textit{Proof of Corollary 1:} This result holds trivially because $(Z \independent X ~|~ f(X)) \Rightarrow (Z \independent X , g(X) ~|~ f(X) )$.\\

The proof of Theorem 1 is broken into two cases. First, we show that the treatment effect is identified given a correctly specified propensity score and any other arbitrary function of the covariates. Second, we show that the treatment effect is identified given correctly specified prognostic scores and any other arbitrary function of the covariates. \\

\textit{Case 1: Assume that the propensity score $\varphi(X)$ is correctly specified, but the prognostic score $h(X)$ is misspecified. Then,  $(Y(1),Y(0) \independent W ~\vert~ \varphi(X) , h(X))$.} \\

This result follow directly from the results of \citep{rosenbaum1983central}. Specifically, $b(X)=(\varphi(X),h(X))$ is trivially a balancing score by Theorem 2 of \citet{rosenbaum1983central}. This balancing score property of $b(X)$ can also be seen in Corollary 1 by taking $Z=W$ and $f(X)=\varphi(X)$. In either case, it is obvious that $b(X)=(\varphi(X),h(X))$ is a balancing score. By Theorem 3 of \citet{rosenbaum1983central}, if the treatment assignment is strongly ignorable given $X$, then it is strongly ignorable given any balancing score. \\ 

\textit{Case 2: Assume that the prognostic scores $\Psi_0(x)$ and $\Psi_1(X)$ are correctly specified, but the propensity score $h(X)$ is misspecified.} \\

First, note that $(Y(0) \independent X | \Psi_0(X) , q(X) )$, where $q(X)=(\Psi_1(X) , h(X))$. This follows directly from Corollary 1 with $Z=Y(0)$ and $f(X)=\Psi_0(X)$. Given this result, we assert that:
\[ Y(0) \independent (W,X) ~|~ \Psi_0(X) , q(X).\]
Proof of this result follows.
\begin{align*}
	P( Y(0) , X , W | \Psi_0(X) , q(X) ) &= P(Y(0)  | W, X, \Psi_0(X),q(X) ) P( W,X|\Psi_0(X),q(X) ) \\
	&= P(Y(0) | X, \Psi_0(X),q(X)) P( W,X|\Psi_0(X),q(X) ) \\&\qquad {\rm{~~~since~}}Y(0) \independent W ~|~ X \\
	&= P(Y(0)| \Psi_0(X),q(X)) P( W,X|\Psi_0(X),q(X) ) \\&\qquad{\rm{~~~since~}} Y(0) \independent X | \Psi_0(X) , q(X)
\end{align*}

\noindent A similar line of arguments can be used for $Y(1)$, but reversing the role of $\Psi_0(X)$ and $\Psi_1(X)$.

\section{Proof of Theorem 2}

\subsection{Proof of Theorem 2}
We provide a proof of Theorem 2 under no effect modification. In the presence of effect modification, we can condition on two prognostics scores in addition to the propensity score to achieve double robustness, or we can target the average treatment effect on the treated. See the main text for further discussion. In either case, the proof follows the same steps as shown here, but with a slight modification to the rates if matching on more than two scores. 

Throughout this section, we assume that the matching scores are estimated on a sample that is independent of the sample used for estimation. This is similar to a discussion that can be found in \citep{hansen2008prognostic}, where it is argued that the same data cannot be used for estimation of both the prognostic score and the treatment effect.

The proof of Theorem 2 is aided by the introduction of a smoothed version of the matching estimator. We show that the difference between the matching estimator and its smoothed version converges to zero and that the smoothed version of the matching estimator is consistent for the effect of interest. We first lay out two useful results that will be used in the proof.

{\bf Result 1:}
Let $U_1,...,U_n$ be an i.i.d. sample with cumulative distribution function and probability distribution function denoted by $F$ and $f$. Then, the probability distribution function of consecutive order statistics is given by:
\begin{align*}
	f_{U_{(m)},U_{(m+1)}}(x,y) &= 
    b_{nm}F(x)^{m-1} \left( 1 - F(y) \right)^{n-m-1} f(x)f(y)~~~,~~~x<y
\end{align*}
for $b_{nm}=\frac{n!}{(m-1)!(n-m-1)!}$.

{\bf Result 2:} Let $U_1,...,U_n$ be an i.i.d. sample with cumulative distribution function and probability distribution function denoted by $F$ and $f$. Then, the cumulative distribution function of the difference of consecutive order statistics is bounded by:
\begin{align*}
	F_{U_{(m+1)}-U_{(m)}}(u) &\leq \int_{-\infty}^{\infty} b_{nm} f(x) \left[ F(x+u)-F(x) \right] \partial x 
\end{align*}

{Proof:}
\begin{align*}
	F_{U_{(m+1)}-U_{(m)}}(u) &= Pr(U_{(m+1)}-U_{(m)} \leq u)\\
    &= \int_{-\infty}^{\infty} \int_{x}^{x+u} f_{U_{(m)},U_{(m+1)}}(x,y) \partial y \partial x \\
    &= \int_{-\infty}^{\infty} b_{nm} F(x)^{m-1} f(x) \left[ \int_{x}^{x+u} \left( 1 - F(y) \right)^{n-m-1} f(y) \partial y \right] \partial x \\
    &\leq \int_{-\infty}^{\infty} b_{nm} f(x) \left[ \int_{x}^{x+u} f(y) \partial y \right] \partial x \\
    &= \int_{-\infty}^{\infty} b_{nm} f(x) \left[ F(x+u)-F(x) \right] \partial x 
\end{align*}

\noindent {\bf Assumptions of Theorem 2:} 

Define $C_{iM}(\theta) = \frac{D_{i(M)}(\theta) + D_{i(M+1)}(\theta)}{2}$, where $D_{i(k)}(\theta)$ indicates the $k$-th order statistic of $\{ D_{ij}(\theta) = \|Z_j(\theta) - Z_i(\theta)\|^2 : ~W_j = 1 - W_i\}$ for a given $i$, and let $D_{ij}(\theta)$ have the density $f_{D;i, \theta}$. Further, let $\ell_{ij}(\theta) = C_{iM}(\theta) - \|Z_j(\theta) - Z_i(\theta)\|^2$. The quantity $\widetilde{\theta} = (\gamma^*_w, \gamma^*_y)$ denotes the probability limit of $\widehat{\theta}$. And, finally, $H_{ij}(\widetilde{\theta}) = Y^2_jY^2_i\left\{\frac{\partial}{\partial \widetilde{\theta}}\ell_{ij}(\widetilde{\theta})\right\}^2\left\{\frac{\partial}{\partial \widetilde{\theta}}\ell_{ji}(\widetilde{\theta})\right\}^2$ and $H_{ijk}(\widetilde{\theta}) = |Y_j||Y_i||Y_k|\left|\frac{\partial}{\partial \widetilde{\theta}}\ell_{ij}(\widetilde{\theta})\right|\left|\frac{\partial}{\partial \widetilde{\theta}}\ell_{ji}(\widetilde{\theta})\right|\left|\frac{\partial}{\partial \widetilde{\theta}}\ell_{ik}(\widetilde{\theta})\right|\left|\frac{\partial}{\partial \widetilde{\theta}}\ell_{ki}(\widetilde{\theta})\right|\left|\frac{\partial}{\partial \widetilde{\theta}}\ell_{kj}(\widetilde{\theta})\right|\left|\frac{\partial}{\partial \widetilde{\theta}}\ell_{jk}(\widetilde{\theta})\right|$.

\begin{enumerate}
	\item SUTVA, strong ignorability, and positivity.
    \item No effect modification.
    \item Regularity conditions necessary for asymptotic consistency of the lasso found in \citep{van2008high, negahban2009unified}. 
    \item Sparsity as defined in Section 2.3 of the main text.
    \item At least 1 of the 2 high dimensional models is correctly specified. That is, either $\varphi(X)=P(W=1|X) = g(X'\gamma^*_w)$ or $\Psi(X)=E(Y|W=0,X) = f(X'\gamma^*_y)$.
    \item The distributions of the matching discrepancies for both known matching scores and estimated matching scores are continuous with bounded second moments of the underlying probability distribution functions. That is, $\int f^2_{D;i, \widetilde{\theta}}(x) dx < \infty$ and $\int f^2_{D;i, \widehat{\theta}}(x) dx < \infty$.
    \item 
$E\left[H_{ij}(\widetilde{\theta})\right] < \infty$ and $E\left[H_{ijk}(\widetilde{\theta})\right] < \infty$.
\end{enumerate}

\noindent {\bf Proof of Theorem 2:}
First, note that we can write the error of our estimator as:

\begin{align*}
	\tau(\widehat{\theta}) - \tau &= \left[\tau(\widehat{\theta}) - \tau(\widetilde{\theta})\right] + \left[ \tau(\widetilde{\theta}) - \tau\right].
\end{align*}
The first component is further decomposed as:
\begin{align*}
\tau(\widehat{\theta}) - \tau(\widetilde{\theta}) &= \left[ \tau(\widehat{\theta}) - \tau_\Phi(\widehat{\theta}; h_N) \right] + \left[ \tau_\Phi(\widehat{\theta}; h_N)  - \tau_\Phi(\widetilde{\theta}; h_N)\right] + \left[\tau_\Phi(\widetilde{\theta}; h_N) - \tau(\widetilde{\theta}) \right]
\end{align*}
where $\tau_\Phi(\theta; h)$ is a smoothed matching estimator defined below. We examine the convergence of each of the four components of this decomposition separately. Specifically, we show that for a properly chosen bandwidth $h_N$ and any $Q\geq0$:

\begin{enumerate}
	\item $\tau(\widehat{\theta}) - \tau_\Phi(\widehat{\theta}; h_N) = o_p\left(N^{-Q}\right)$
    \item $\tau_\Phi(\widetilde{\theta}; h_N) - \tau(\widetilde{\theta})= o_p\left(N^{-Q}\right)$
    \item $\tau_\Phi(\widehat{\theta}; h_N)  - \tau_\Phi(\widetilde{\theta}; h_N) = o_p\left(\sqrt{\frac{\log P}{N}}\right)$
    \item $\tau(\widetilde{\theta}) - \tau = O_p\left(N^{-1/2}\right)$
\end{enumerate}

\noindent Combining the rates of convergence leads directly to $\tau(\widehat{\theta}) - \tau = O_p\left(\sqrt{\frac{\log P}{N}}\right)$. Note that \#1 and \#2 indicate that the difference between the matching estimator and its smoothed version converges at a rate that is faster than any polynomial. This was achieved by choosing a bandwidth $h_N$ that converges to 0 suitably fast, thus ensuring the difference between the matching estimator and its smoothed version is small.  

{\bf Proof of \#1 and \#2}


First, we rewrite $\tau(\theta)$:
\def\thetatilde{\widetilde{\theta}}
\def\Xtilde{\widetilde{X}}
\begin{align*}
\tau(\theta) 
&= \frac{1}{N}\sum_{i=1}^N(2W_i - 1)\left[Y_i-\frac{1}{M}\sum_{j: W_j = 1 - W_i} I\left\{C_{iM} - \|Z_j(\theta) - Z_i(\theta)\|^2 > 0 \right\}Y_j\right]
\end{align*}
for $C_{iM} = \frac{D_{i(M)} + D_{i(M+1)}}{2}$, where $D_{i(k)}$ indicates the $k$-th order statistic of $\{ D_{ij} = \|Z_j(\theta) - Z_i(\theta)\|^2 : ~W_j = 1 - W_i\}$ for a given $i$. 

Now, define a smooth version of $\tau(\theta)$ where the indicator is replaced by a smooth function \[
\tau_\Phi(\theta; h_N) = \frac{1}{N}\sum_{i=1}^N(2W_i - 1)\left[Y_i-\frac{1}{M}\sum_{j: W_j = 1 - W_i} \Phi_{h_N}\left\{\ell_{ij}(\theta)\right\}Y_j\right]
\]
where $\ell_{ij}(\theta) = C_{iM} - \|Z_j(\theta) - Z_i(\theta)\|^2$ and $\Phi_{h_N}(x) = (1 + e^{-x/h_N})^{-1}$. 

For any $\theta$ then, the difference $\tau_\Phi(\theta; h_N) - \tau(\theta)$ is given by:
\begin{align*}
&\left|\tau_\Phi(\theta; h_N) - \tau(\theta)\right| \\
&=\left|\frac{1}{NM}\sum_{i=1}^N\sum_{j: W_j = 1 - W_i}(2W_i - 1)Y_j\left[\Phi_{h_N}\left\{\ell_{ij}(\theta)\right\} - I\left\{\ell_{ij}(\theta) > 0\right\} \right]\right|\\
&\leq \frac{1}{M}\sqrt{\frac{1}{N}\sum_{j: W_j = 1 - W_i}Y_j^2}\sqrt{\sum_{i=1}^N\sum_{j: W_j = 1 - W_i}\left[\Phi_{h_N}\left\{\ell_{ij}(\theta)\right\} - I\left\{\ell_{ij}(\theta) > 0\right\} \right]^2}\\
&= O_p(1)\sqrt{\sum_{i=1}^N\sum_{j: W_j = 1 - W_i}\left[\Phi_{h_N}\left\{\ell_{ij}(\theta)\right\} - I\left\{\ell_{ij}(\theta) > 0\right\} \right]^2}.
\end{align*}

The difference $\Phi_{h_N}\left\{\ell_{ij}(\theta)\right\} - I\left\{\ell_{ij}(\theta) > 0\right\}$ can be written as:
\begin{align*}
\Phi_{h_N}\left\{\ell_{ij}(\theta)\right\} - I\left\{\ell_{ij}(\theta) > 0\right\} = 
 \frac{\text{sign}\{-\ell_{ij}(\theta)\}}{e^{|\ell_{ij}(\theta)|/h_N} + 1}.
\end{align*}
\def\thetahat{\widehat{\theta}}
\def\lll{\mathcal{L}}
\noindent Let $\lll_i(\theta)=\min_{\{j:W_j=1-W_i\}}(\ell_{ij}(\theta)) = \frac{D_{i(M+1)}-D_{i(M)}}{2}$. Note that for a given $Z_i(\theta)$, $D_{i(M)}$ are similar to the matching discrepancies discussed in \cite{abadie2006large}, but their results are not directly applicable here because we are concerned with the difference in consecutive discrepancies as define by $\lll_i(\theta)$. 

Now, choose a bandwidth that converges to 0 sufficiently fast. One such choice is $h_N = \frac{1}{N^3b_{NM}}$. Taking $\theta = \widehat{\theta}$ we can write the following:

\begin{align*}
\left|\tau_\Phi(\thetahat; h_N) - \tau(\thetahat)\right| &\leq O_p(1)\sqrt{\sum_{i=1}^N\sum_{j: W_j = 1 - W_i}\frac{1}{\left(e^{|\ell_{ij}(\thetahat)|/h_N} + 1\right)^2}} \\
&\leq O_p(1)\sqrt{\sum_{i=1}^N\frac{N}{e^{\lll_i(\thetahat)/h_N} }} 
\end{align*}

Now, for any $0<K\leq1$, and $Q \geq 0$: 
\def\inv{^{-1}}
{\footnotesize
\begin{align*}
\lim_{N\rightarrow \infty}  Pr\left\{\sum_{i=1}^N\frac{N^{Q+1}}{e^{\lll_i(\thetahat)/h_N}} > K \right\} &\leq \lim_{N\rightarrow \infty} \sum_{i=1}^N Pr\left\{\frac{N^{Q+1}}{e^{\lll_i(\thetahat)/h_N}} > \frac{K}{N} \right\} \\
&\leq \lim_{N\rightarrow \infty} \sum_{i=1}^N Pr\left\{\lll_i(\thetahat) < \frac{-\log K+(Q+2)\log N}{N^3b_{NM}} \right\} \\
&\leq \lim_{N\rightarrow \infty} \sum_{i=1}^N Pr\left\{ \frac{D_{i(M+1)}(\thetahat)-D_{i(M)}(\thetahat)}{2} < \frac{-\log K+(Q+2)\log N}{N^3b_{NM}} \right\} \\
&\leq \lim_{N\rightarrow \infty} \sum_{i=1}^N \int_{-\infty}^{\infty} b_{NM}  f_i(x) \left[ F_i\left(x+\frac{-2\log K+2(Q+2)\log N}{N^3b_{NM}} \right) - F_i(x) \right] \partial x \\
\end{align*}
where $F_i = F_{D;i, \widehat{\theta}}$ and $f_i = f_{D;i, \widehat{\theta}}$ are the cumulative distribution function and probability distribution functions of $\{ D_{ij}(\thetahat) = \|Z_j(\thetahat) - Z_i(\thetahat)\|^2 : ~W_j = 1 - W_i\}$ for a given $i$. Note that these $D_{ij}(\thetahat)$ for a given $i$ are independent since $\thetahat$ is estimated on a sample that is independent from the estimation sample. Next, we expand $F_i\left(x+u\right)$ around $x$ such that $F_i\left(x+u\right) = F_i\left(x\right) + u f_i(x^*)$ for some $x^* \in [x,x+u]$

\begin{align*}
\lim_{N\rightarrow \infty}  Pr\left\{\sum_{i=1}^N\frac{N^{Q+1}}{e^{\lll_i(\thetahat)/h_N}} > K \right\} &\leq \lim_{N\rightarrow \infty} \sum_{i=1}^N \int_{-\infty}^{\infty} b_{NM} f_i\left(x\right) \left[ \frac{-2\log K+2(Q+2)\log N}{N^3b_{NM}} f_i\left(x^*\right) \right] \partial x \\
&\leq \lim_{N\rightarrow \infty} \sum_{i=1}^N  \frac{-2\log K+2(Q+2)\log N}{N^3}  \int_{-\infty}^{\infty}f^2_i\left(x\right)\partial x \\
&\leq C \lim_{N\rightarrow \infty} \frac{-2N\log K+2(Q+2)N\log N}{N^3} \\
&\leq 0\\
\end{align*}
}
since $\int_{-\infty}^{\infty}f^2_i\left(x\right)\partial x$ is bounded for all $i$ by assumption. This shows that 
\def\thetatilde{\widetilde{\theta}}
\begin{align*}
\left|\tau_\Phi(\thetahat; h_N) - \tau(\thetahat)\right| &\leq O_p(1) o_p(N^{-Q}) \\
&= o_p(N^{-Q}),~~~\rm{for~any}~Q\geq 0
\end{align*}
and an identical argument ensures  $\left|\tau_\Phi(\thetatilde; h_N) - \tau(\thetatilde)\right| = o_p(N^{-Q})$. 

{\bf Proof of \#3}

We may now analyze the smoothed version of $\tau(\cdot)$ to determine its behavior as $\widehat{\theta}$ converges to $\widetilde{\theta}$. First note that a Taylor expansion yields
\begin{align*}
	\tau_\Phi(\widehat{\theta}; h_N) - \tau_\Phi(\widetilde{\theta}; h_N)
	&=  \frac{\partial }{\partial \widetilde{\theta}}\tau_\Phi(\widetilde{\theta};h_N)'(\widehat{\theta} - \widetilde{\theta}) + O_p\left(\|\widehat{\theta} - \widetilde{\theta}\|^2\right)
\end{align*}
 and
\def\bone{{\bf 1}}

\begin{align*}
\left|\frac{\partial }{\partial \widetilde{\theta}}\tau_\Phi(\widetilde{\theta};h_N)\right| &= \left|\left(\frac{\partial}{\partial \widetilde{\theta}} \frac{1}{N}\sum_{i=1}^N(2W_i - 1)\left[Y_i-\frac{1}{M}\sum_{j: W_j = 1 - W_i} \Phi_{h_N}\left\{\ell_{ij}(\widetilde{\theta}) \right\}Y_j\right]\right)\right|\\
&= \left|\frac{2}{NM}\sum_{i=1}^N(2W_i - 1)\sum_{j: W_j = 1 - W_i}Y_j\phi_{h_N}\left\{\ell_{ij}(\widetilde{\theta}) \right\}\frac{\partial}{\partial \widetilde{\theta}}\ell_{ij}(\widetilde{\theta})\right|\\
&\leq \left|\frac{2}{NM}\sum_{i=1}^N\phi_{h_N}\left\{\lll_i(\widetilde{\theta}) \right\} \sum_{j: W_j = 1 - W_i}Y_j\frac{\partial}{\partial \widetilde{\theta}}\ell_{ij}(\widetilde{\theta})\right|\\
&\leq \frac{2}{NM}\sum_{i=1}^N\phi_{h_N}\left\{\lll_i(\widetilde{\theta}) \right\} \sum_{j: W_j = 1 - W_i}\left|Y_j\frac{\partial}{\partial \widetilde{\theta}}\ell_{ij}(\widetilde{\theta})\right|\\
&\leq \frac{2}{NM}\sqrt{\sum_{i=1}^N\phi^2_{h_N}\left\{\lll_i(\widetilde{\theta}) \right\}}\sqrt{\sum_{i=1}^N\left(\sum_{j: W_j = 1 - W_i}\left|Y_j\frac{\partial}{\partial \widetilde{\theta}}\ell_{ij}(\widetilde{\theta})\right|\right)^2}\\
&\leq \frac{2}{M}\sqrt{N\sum_{i=1}^N\phi^2_{h_N}\left\{\lll_i(\widetilde{\theta}) \right\}}\sqrt{N^{-3}\left\{\sum_{i \neq j} H_{ij}(\widetilde{\theta}) + \sum_{i \neq j \neq k} H_{ijk}(\widetilde{\theta})\right\}}
\end{align*}
where $\phi_h(x) =  \partial\Phi_h(x)/\partial x = \frac{1}{h} \frac{e^{-x/h}}{\left(1+e^{-x/h}\right)^2}$. 

Under Assumption 7,  $\frac{1}{N(N-1)}\sum_{i\neq j}H_{ij}(\widetilde{\theta})$ and $\frac{1}{N(N-1)(N-2)}\sum_{i\neq j}H_{ijk}(\widetilde{\theta})$ are each $O_p(1)$ because of the law of large numbers for U-statistics \cite{hoeffding1961strong}, and thus $N^{-3}\left\{\sum_{i \neq j} H_{ij}(\widetilde{\theta}) + \sum_{i \neq j \neq k} H_{ijk}(\widetilde{\theta})\right\} = O_p(1)$. Now, for any $0< K \leq 1$, a similar line of arguments as in the proof of \#1 and \#2 shows that:

\begin{align*}
 \lim_{N \rightarrow \infty} Pr\left\{N\sum_{i=1}^{N}\phi^2_{h_N}(\lll_i(\widetilde{\theta}) ) > K\right\} &\rightarrow 0 \\
\end{align*}
Therefore,

\begin{align*}
\left|\frac{\partial }{\partial \widetilde{\theta}}\tau_\Phi(\widetilde{\theta};h)\right|&= o_p(1)
\end{align*}

The convergence of $\tau_\Phi(\widehat{\theta}; h_N) - \tau_\Phi(\widetilde{\theta}; h_N)$, combined with the $\sqrt{\frac{\log P}{N}}$ rate of convergence of $\hat{\theta}$ to $\tilde{\theta}$ available in, e.g., \cite{van2008high} follows: 

\begin{align}
\tau_\Phi\left(\widehat{\theta}; h_N\right) - \tau_\Phi\left(\widetilde{\theta}; h_N\right) = o_p\left(\sqrt{\frac{\log P}{N}}\right)\label{tau_phi}
\end{align}


{\bf Proof of \#4}

Assuming SUTVA, strong ignorability, and positivity hold, and at least one of the two score models is correctly specified, then $$\tau(\widetilde{\theta}) - \tau = O_p(N^{-1/2}).$$

This follows from Theorem 1 of \cite{abadie2006large}, which ensures that $$
\frac{1}{N} \sum_{i=1}^N (2W_i - 1) \left( Y_i - \frac{1}{M} \sum_{j \in \mathcal{J}^*_M(i, U)}Y_j \right) - E\left[E\left\{Y \left| U, W = 1\right.\right\} - E\left\{Y \left| U, W = 0\right.\right\}\right] = O_p(N^{-1/k})
$$
where the  matching set $\mathcal{J}^*_M(i, U)$ is constructed based on $U$, a vector of $k$ fixed matching variables, $k \geq 2$. Let $U = \left\{f(X'\gamma^*_y), g(X'\gamma^*_w)\right\}$ and note that $k=2$. Theorem 1 of this paper ensures that:

\begin{align*}
	E\left[E\left\{Y \left| f(X'\gamma^*_y), g(X'\gamma^*_w), W = 1\right.\right\} - E\left\{Y \left| f(X'\gamma^*_y), g(X'\gamma^*_w), W = 0\right.\right\}\right] &= \tau, 
\end{align*}

\noindent provided that at least one of the two score models is correctly specified.




\section{Additional simulation results under model misspecification}

These additional simulations are used to illustrate two goals: To assess the performance of the various estimators when only one model is correctly specified, and to confirm our theoretical results that our estimator is doubly robust. With this goal in mind we run simulations for a grid of sample size and covariate dimensions. Specifically, we will vary the sample size, $n \in \{200,500,1000,2500,5000,10000\}$, and the number of covariates, $p \in \{10,50,100,250,500,1000\}$. When the treatment model is misspecified, we will be generating the treatment from the data generation scenario in Section 3.2 of the manuscript, while the outcome will be generated from the linear scenario of Section 3.1. When the outcome model is misspecified, we will be generating the outcome from the data generation scenario in Section 3.2 of the manuscript, while the treatment will be generated from the linear scenario of Section 3.1. For each value of $p$, we will plot the mean squared error as a function of the sample size for each estimator. For ease of illustration we will restrict attention to the three estimators that are doubly robust: The double matching estimator, the lasso based doubly robust estimator, and the estimator from \cite{farrell2015robust} (Farrell). We also provide the analagous results here when both models are misspecified for comparison.

\begin{figure}[htbp]
\centering
\includegraphics[width=0.9\linewidth]{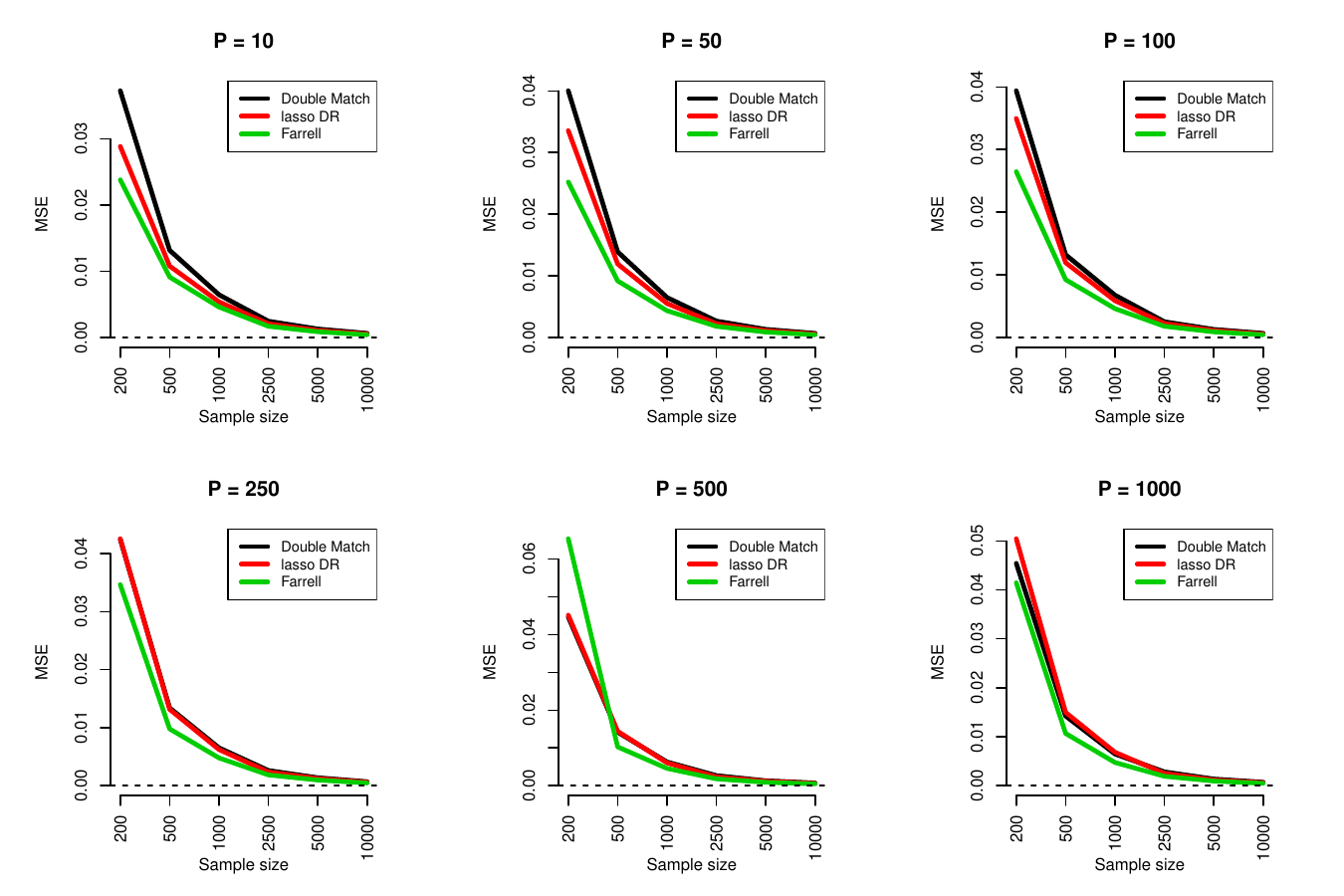}
\caption{Results when the treatment model is misspecified.}
\label{fig:misX}
\end{figure}

\begin{figure}[htbp]
\centering
\includegraphics[width=0.9\linewidth]{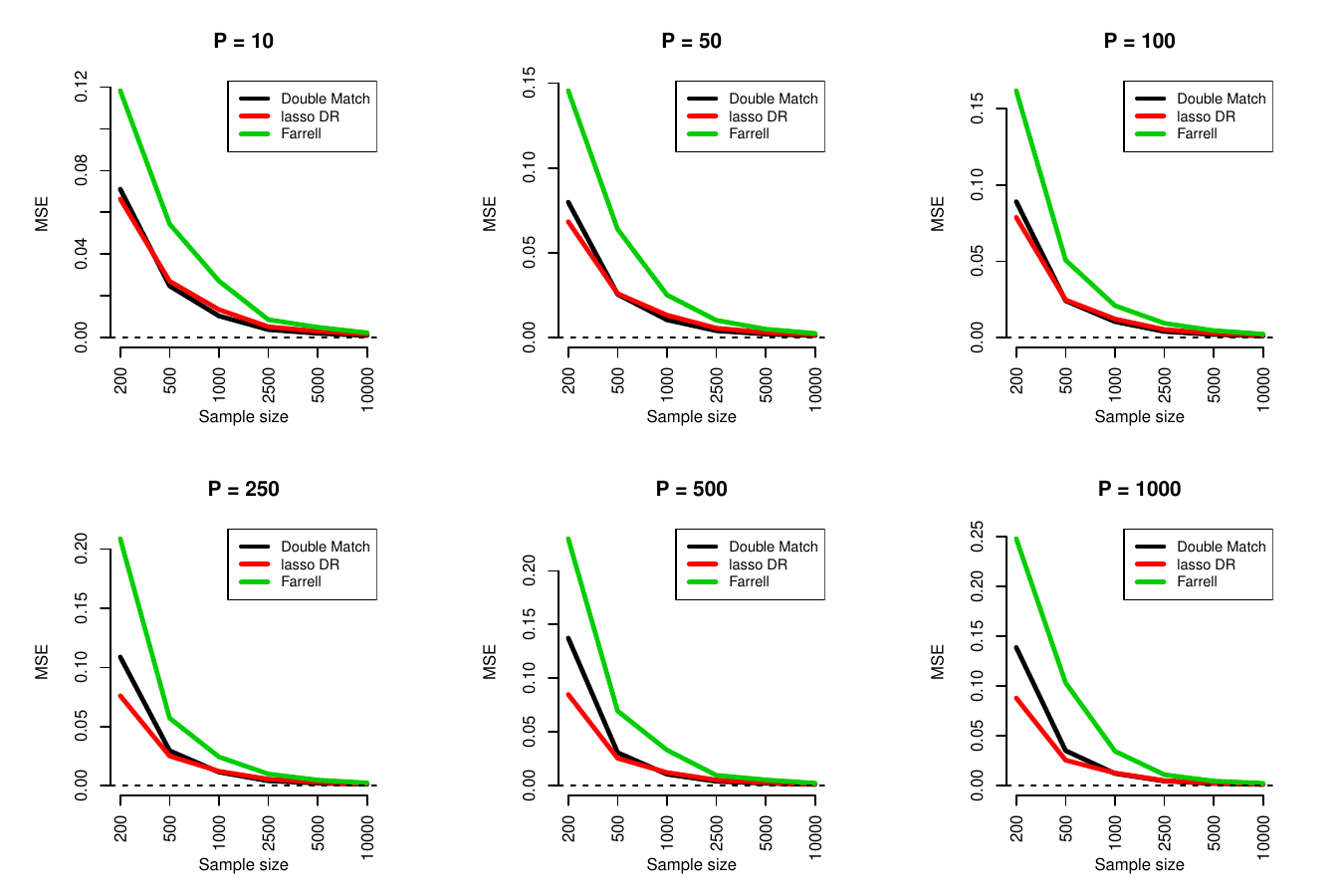}
\caption{Results when the outcome model is misspecified.}
\label{fig:misY}
\end{figure}

\begin{figure}[htbp]
\centering
\includegraphics[width=0.9\linewidth]{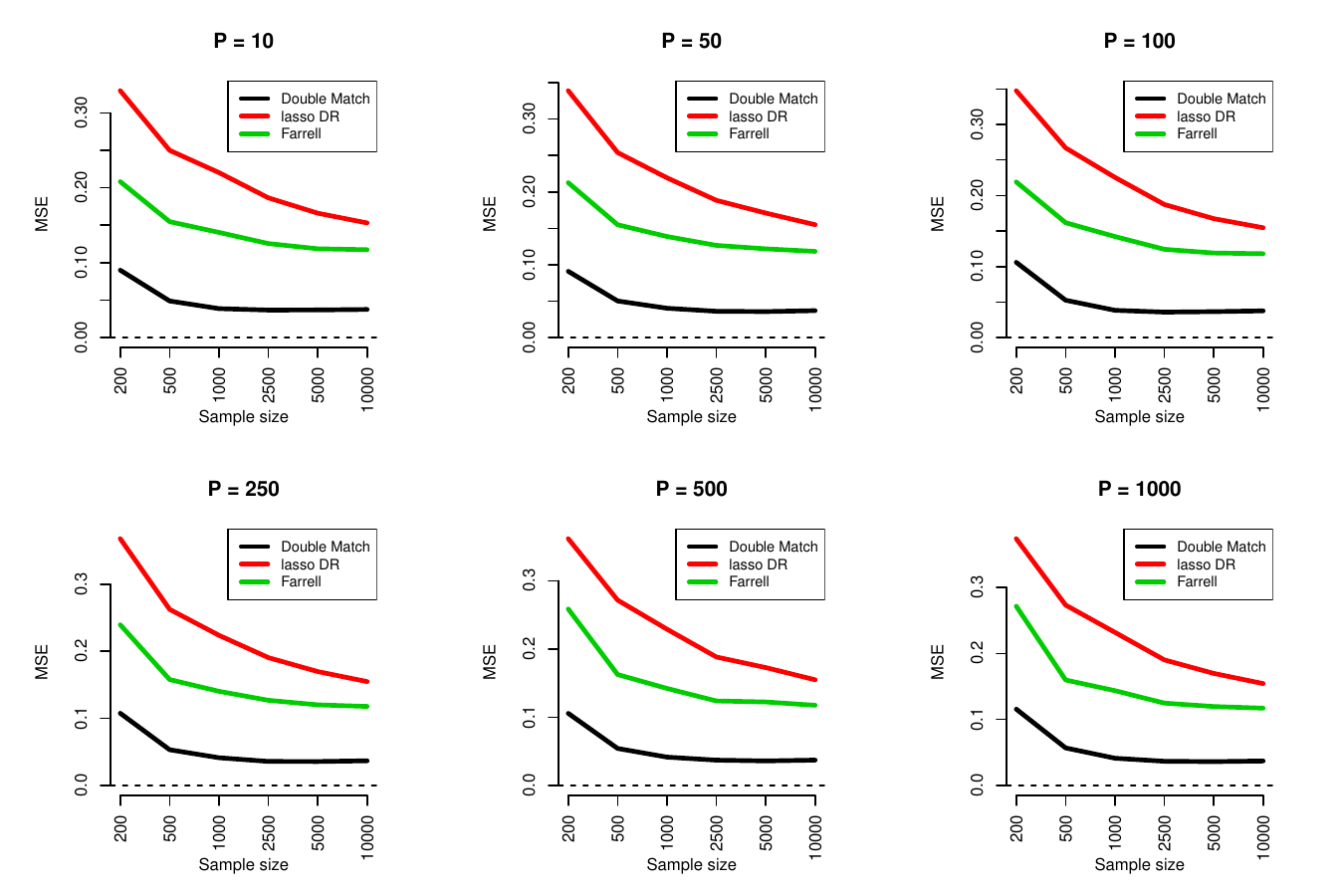}
\caption{Results when both models are misspecified.}
\label{fig:misXY}
\end{figure}

Figure \ref{fig:misX}, Figure \ref{fig:misY}, and Figure \ref{fig:misXY} show the results when the treatment model, outcome model, and both models are misspecified, respectively. The first thing to note, is that as the sample size increases, our estimator has an MSE converging to zero when only one model is misspecified, confirming the double robust property of our estimator. When the treatment model is misspecified, all three estimators perform similarly, with the Farrell estimator performing slightly best in terms of MSE. When the outcome model is misspecified, the double matching and lasso DR estimators perform similarly, with the lasso DR approach slightly better in small samples, while the Farrell estimator does the worst. Finally, when both models are misspecified our double matching estimator drastically outperforms the other estimators in terms of MSE. While none of them converge to zero, as expected due to model misspecification, the matching estimator is the most robust to model misspecification under any combination of $n$ and $p$. 

\section{Confirming rate of convergence}

\begin{figure}[htbp]
\centering
\includegraphics[width=0.9\linewidth]{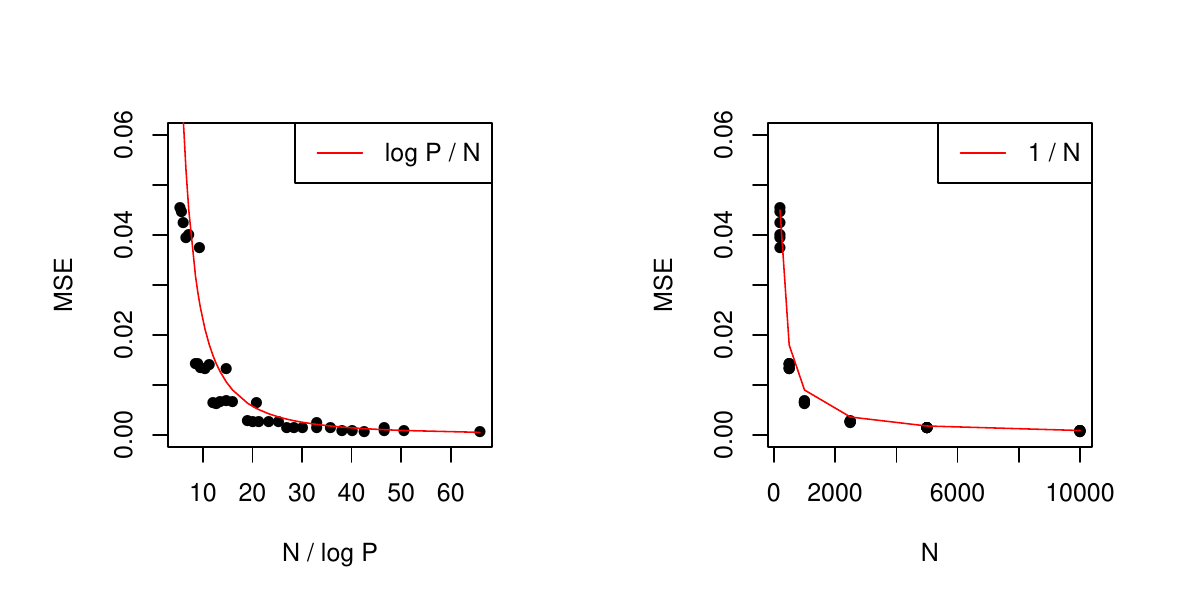}
\caption{Convergence rate when the treatment model is misspecified}
\label{fig:convergenceX}
\end{figure}

\begin{figure}[htbp]
\centering
\includegraphics[width=0.9\linewidth]{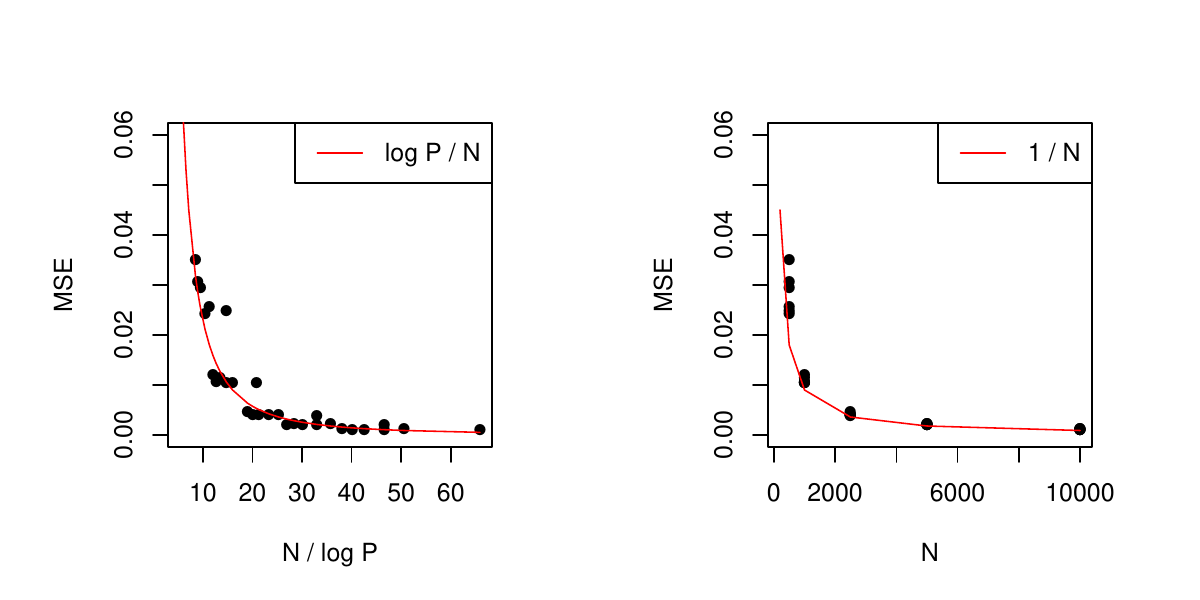}
\caption{Convergence rate when the outcome model is misspecified}
\label{fig:convergenceY}
\end{figure}

We can use the results of the simulations above to empirically verify our theoretical results from the manuscript. In particular, we will attempt to verify that our estimator converges at the $O_p\left(\sqrt{\frac{\log P}{N}}\right)$ rate. To achieve this goal, we will plot the bias of our estimator as a function of $\sqrt{\frac{n}{\log P}}$ and compare it to a line that is proportional to $\sqrt{\frac{\log P}{N}}$. If our estimated bias decreases at similar rate as this line, this would suggest that our estimator has the desired rate. We also plot the bias as a function of $N$ with a line proportional to $1/\sqrt{N}$ to examine if our estimator is approaching the $O_p\left(\sqrt{\frac{1}{N}}\right)$ rate, even though there is no theory to justify this. Figure \ref{fig:convergenceX} shows the results when the treatment model is misspecified and Figure \ref{fig:convergenceY} shows the results when the outcome model is misspecified. We see that in both cases, the estimator's bias decreases at a rate that is faster than $O_p\left(\sqrt{\frac{\log P}{N}}\right)$, and it appears that the rate of convergence is more closely aligned with the $O_p\left(\sqrt{\frac{1}{N}}\right)$ rate. This confirms the theoretical results of our paper and highlights its ability to adjust for confounding in high-dimensional scenarios. 

\section{Additional nonlinear simulation results}

In this section we present results from an additional simulation that looks at the scenario when both models are again misspecified to assess the robustness of our proposed matching procedure against model misspecification. Specifically the data generating mechanism was as follows:
\begin{align}
	W &\sim Bernoulli\left\{\frac{exp(0.7*exp(X_1) + 0.7*log(0.7*X_1^2) - 0.8*X_2^3 + 0.7*X_3^3 - 
              0.5*X_4^3 - 0.8*X_5^2)} {1 + exp(0.7*exp(X_1) + 0.7*log(0.7*X_1^2) - 0.8*X_2^3 + 0.7*X_3^3 - 
              0.5*X_4^3 - 0.8*X_5^2)}\right\} \\
    Y &\sim Normal(-2 + W + 0.7*exp(0.6*X_1) - 0.6*X_2^3 + 0.7*X_3^2, \sigma^2 = 1). 
\end{align}

This is a highly nonlinear situation in which we would expect the linear approximation to the true treatment and outcome models would do quite poorly. In \ref{tab:nonlinear_app} we still see drastic improvements when using the doubly robust matching estimator relative to other approaches that also rely on linear models. 

\begin{table}[hbpt]
\centering
\begin{tabular}{lrrr}
  \hline
type & Absolute bias & SD & MSE \\ 
  \hline 
  Naive & 1.94 & 0.51 & 4.03 \\ 
  Outcome Lasso & 1.45 & 0.50 & 2.35 \\ 
  Double post selection & 0.99 & 0.47 & 1.19 \\ 
  lasso IPW & 1.21 & 0.42 & 1.64 \\ 
  Farrell & 0.69 & 0.86 & 1.22 \\ 
  lasso DR & 1.26 & 0.41 & 1.76 \\ 
  Match PS & 0.66 & 0.92 & 1.27 \\ 
  Match Prog & 0.56 & 0.71 & 0.80 \\ 
  Double Match & 0.13 & 0.54 & 0.30 \\  
   \hline
\end{tabular}
\caption{Nonlinear simulation results.}
\label{tab:nonlinear_app}
\end{table}

\section{Performance of standard error in nonlinear simulation}

In this section we evaluate the use of the standard error estimator in the nonlinear simulation study of Section 3.2. Table \ref{tab:SE} presents the 95\% interval coverages across 1000 simulations in the nonlinear setting for a variety of $N$ and $P$ combinations. It appears that the standard error estimate does a good job in small sample sizes, but does worse as $N$ increases. This is because the doubly robust matching estimator is biased in this scenario under any sample size, and as the sample size grows the standard error tends to zero leading to degraded performance of the confidence intervals.  

\begin{table}[ht]
\centering
\begin{tabular}{ | r | r | r | r | r |}
  \hline
   & P = 200 & P = 500 & P = 1000 & P = 2000 \\ 
  \hline
  N = 200 & 0.97 & 0.97 & 0.97 & 0.96 \\ 
   N = 500 & 0.97 & 0.95 & 0.96 & 0.97 \\ 
   N = 1000 & 0.91 & 0.91 & 0.92 & 0.92 \\ 
   N = 2000 & 0.78 & 0.78 & 0.78 & 0.76 \\  
  \hline
\end{tabular}
\caption{Coverage probabilities for a variety of data dimensions using the proposed standard error estimate in the nonlinear setting of Section 3.2 of the manuscript.}
\label{tab:SE}
\end{table}

\section{Simulation results with different variance values}

In this simulation study we change the strength of confounding by simulating data under the same data generating models as in the main manuscript, however, we will try $\sigma^2 = 0.5$ and $\sigma^2 = 2$. Tables \ref{tab:siglin1} and \ref{tab:siglin2} show the results from the linear simulation scenario from Section 3.1, while Tables \ref{tab:sigmis1} and \ref{tab:sigmis2}. We see that the main conclusions from the paper remain the same. In the linear cases, the doubly robust matching estimator does well in terms of MSE, though is slightly less efficient than the double post selection approach, which uses the correct linear model to estimate the treatment effect. In nonlinear simulation scenarios, the doubly robust matching estimator performs the best in terms of MSE.

\begin{table}[ht]
\centering
\begin{tabular}{lrrr}
  \hline
type & Absolute bias & SD & MSE \\ 
  \hline
True & 0.01 & 0.12 & 0.01 \\ 
  Naive & 0.49 & 0.26 & 0.31 \\ 
  Outcome Lasso & 0.24 & 0.14 & 0.08 \\ 
  Double post selection & 0.06 & 0.14 & 0.02 \\ 
  lasso IPW & 0.36 & 0.21 & 0.17 \\ 
  Farrell & 0.03 & 0.54 & 0.30 \\ 
  lasso DR & 0.19 & 0.13 & 0.05 \\ 
  Match PS & 0.28 & 0.42 & 0.25 \\ 
  Match Prog & 0.21 & 0.17 & 0.07 \\ 
  Double Match & 0.10 & 0.17 & 0.04 \\ 
   \hline
\end{tabular}
\caption{Additional simulation results when the true treatment and outcome models are linear, and $\sigma^2 = 0.5$}
\label{tab:siglin1}
\end{table}

\begin{table}[ht]
\centering
\begin{tabular}{lrrr}
  \hline
type & Absolute bias & SD & MSE \\ 
  \hline
True & 0.00 & 0.21 & 0.04 \\ 
  Naive & 0.48 & 0.31 & 0.32 \\ 
  Outcome Lasso & 0.34 & 0.22 & 0.17 \\ 
  Double post selection & 0.09 & 0.25 & 0.07 \\ 
  lasso IPW & 0.36 & 0.26 & 0.20 \\ 
  Farrell & 0.10 & 0.39 & 0.16 \\ 
  lasso DR & 0.26 & 0.21 & 0.11 \\ 
  Match PS & 0.30 & 0.48 & 0.32 \\ 
  Match Prog & 0.27 & 0.35 & 0.19 \\ 
  Double Match & 0.07 & 0.33 & 0.11 \\  
   \hline
\end{tabular}
\caption{Additional simulation results when the true treatment and outcome models are linear, and $\sigma^2 = 2$}
\label{tab:siglin2}
\end{table}

\begin{table}[ht]
\centering
\begin{tabular}{lrrr}
  \hline
type & Absolute bias & SD & MSE \\ 
  \hline
  Naive & 0.58 & 0.28 & 0.41 \\ 
  Outcome Lasso & 0.57 & 0.25 & 0.38 \\ 
  Double post selection & 0.33 & 0.25 & 0.17 \\ 
  lasso IPW & 0.54 & 0.27 & 0.36 \\ 
  Farrell & 0.38 & 0.28 & 0.22 \\ 
  lasso DR & 0.52 & 0.25 & 0.33 \\ 
  Match PS & 0.35 & 0.40 & 0.28 \\ 
  Match Prog & 0.23 & 0.20 & 0.09 \\ 
  Double Match & 0.09 & 0.26 & 0.07 \\    \hline
\end{tabular}
\caption{Additional simulation results when the true treatment and outcome models are nonlinear, and $\sigma^2 = 0.5$}
\label{tab:sigmis1}
\end{table}

\begin{table}[ht]
\centering
\begin{tabular}{lrrr}
  \hline
type & Absolute bias & SD & MSE \\ 
  \hline
  Naive & 0.61 & 0.33 & 0.48 \\ 
  Outcome Lasso & 0.61 & 0.31 & 0.47 \\ 
  Double post selection & 0.38 & 0.35 & 0.27 \\ 
  lasso IPW & 0.57 & 0.32 & 0.43 \\ 
  Farrell & 0.44 & 0.44 & 0.39 \\ 
  lasso DR & 0.56 & 0.32 & 0.41 \\ 
  Match PS & 0.40 & 0.53 & 0.44 \\ 
  Match Prog & 0.28 & 0.34 & 0.19 \\ 
  Double Match & 0.03 & 0.38 & 0.15 \\ 
   \hline
\end{tabular}
\caption{Additional simulation results when the true treatment and outcome models are nonlinear, and $\sigma^2 = 2$}
\label{tab:sigmis2}
\end{table}

\bibliographystyle{authordate1}
\bibliography{DoubleMatching}

\end{document}